# The Registry of Scientometric Data Sources

Grischa Fraumann[1], Svantje Lilienthal[1], & Christian Hauschke[1]

[1]Research & Development Department, TIB – Leibniz Information Centre for Science and Technology, Welfengarten 1 B, 30167 Hannover, Germany

## ABSTRACT

In this article, we describe the *Registry of Scientometric Data Sources (RSDS)* and several scientometric data sources recorded in this open registry that could be of interest for scientometricians, institutional researchers, librarians, practitioners, policy makers, students and other stakeholders with an interest in scientometrics. This registry was created after carrying out a literature review and a technical evaluation of several data sources. Each data source is recorded with descriptive metadata fields and URLs to further information. This article describes the motivation behind the development of the registry, explains the features that are available on its public website (https://labs.tib.eu/rosi), and closes with a call for participation.

**Corresponding Authors:**
Grischa Fraumann
gfr@hum.ku.dk; https://orcid.org/0000-0003-0099-6509
Svantje Lilienthal
svantje.lilienthal@wikimedia.de, https://orcid.org/0000-0003-1537-2862
Christian Hauschke
christian.hauschke@tib.eu; http://orcid.org/0000-0003-2499-7741

## 1. INTRODUCTION

The number and variety of scientometric online data sources are constantly increasing, which is a stark contrast to the handful of sources that existed a few years ago. These more traditional sources are widely discussed in the literature, such as the bibliographic databases *Scopus* (Baas et al., 2020), and *Web of Science (WoS)* (Birkle et al., 2020), and the academic search engine *Google Scholar* (Gusenbauer, 2019). In the digital era, there is a need for a comprehensive overview on conventional and emerging (open data) sources (Waltman & Larivière, 2020), which was expressed in several national and international initiatives, such as the Leiden Manifesto for Research Metrics (Hicks et al., 2015).

We describe an open collaborative public registry of such sources, the *Registry of Scientometric Data Sources (RSDS)*, which provides such an overview and to which users from the scholarly community and beyond are invited to contribute (Lilienthal, 2019). For that, we formulated the following research question: *Which scientometric data sources are available for research purposes?* First, we evaluated the data sources internally as part of the research project "Reference Implementation for Open Scientometric Indicators" (ROSI) (Hauschke et al., 2018) and later decided to share our findings in the form of the RSDS[1]. The registry is a service to scientometricians, institutional researchers, librarians, practitioners, policy makers, students and other stakeholders with an interest in scientometrics who would like to decide which data sources to use in research projects, to search for or add new data sources. We adopted a crowdsourcing approach that had also been used in similar initiatives (Heibi et al., 2019; OSM Consortium, 2018). The registry is, thus, a new open research infrastructure. Along with open access to publications and research data, research infrastructures are being opened more and more to the wider society. For example, the JRC (Joint Research Centre) Research Infrastructures by the European Commission were opened via calls for participation to the scholarly community and the wider public (European Commission, 2020).

## 1.1 Background and Literature Review

The rise of scientometrics is closely linked to the availability of machine-readable data sources. In the early days of the discipline, it was required to manually aggregate data from the publications themselves or from bibliographies. A prominent example of the latter is the work done by Lotka (1926) for "The frequency distribution of scientific productivity", where he made use of the index of persons in the Chemical Abstracts 1907–1916 and in Auerbach's

[1] https://labs.tib.eu/rosi

*Geschichtstafeln der Physik* to identify prominent researchers according to the number of publications. Although his method led to success, it was naturally very tedious.

The first tool to provide convenient electronic access to citation information was the *Science Citation Index (SCI)* which was introduced in 1955 in the journal *Science* and released in 1964 (Garfield, 2007). Additionally, the *SCI Journal Citation Reports (JCR)* were launched in 1975 and the impact factor rankings were established. Put simply, the impact factor calculates the average number of citations per paper of a given journal, usually based on the last two years. The impact factor has received a considerable amount of criticism which is not the focus of this article (Lariviere & Sugimoto, 2019). The SCI displays the number of publications (i.e., the *Source Author Index*) and the citations of individual researchers (i.e., the *Citation Index*). In the JCR, the citations are ordered according to the journal name. These two measures can also be extended to the institutional and/or country level. The bibliographic database *Web of Science (WoS)* was described by Garfield (2007) as the electronic version of the SCI, among others.

The next evolutionary step was the availability of scientometric data via application programming interfaces (APIs). Put simply, an API "is an interface that provides programmatic access to service functionality and data within an application or a database" (Gartner, 2020). APIs become more and more common in distributed research environments. For example, an API allows researchers to retrieve always the latest data from an interface, and it makes it possible to integrate it into another system (National Information Standards Organization (NISO), 2016). The availability of APIs is also an approach that can be studied when investigating data sources. Another aspect is the availability of documentations that describe the use of APIs (National Information Standards Organization (NISO), 2016). APIs are widely used for research purposes; even systems that usually require a subscription, such as *Scopus*, provide such a service to researchers or other users (National Information Standards

Organization (NISO), 2016). Obviously, public services, such as Wikipedia, also provide APIs to retrieve data (National Information Standards Organization (NISO), 2016). However, some well-known academic search engines, such as *Google Scholar* do not provide access via APIs (Balstad & Berg, 2020). Also, there is a risk that some APIs could get discontinued. For example, the API by the professional networking site *LinkedIn* had been used since 2013 for altmetric studies (Altmetric.com, 2021a), for instance, to investigate how research outputs were shared on public profiles and in groups. However, the API was discontinued in 2014. Some data providers offer a push API which enables data sources to push data live into the system or into several instances (National Information Standards Organization (NISO), 2016). The currentness of API responses may depend on the data provider (National Information Standards Organization (NISO), 2016). Some of these providers provide a so-called premium access through a service-level agreement which guarantees certain functions, such as response times or number of calls in a specified time range (National Information Standards Organization (NISO), 2016). Such a limitation was introduced, for instance, by *Crossref Event Data (CED)* (Crossref Event Data, 2021), a service by the publisher consortium Crossref that tracks various online events, such as the mentions of research outputs in social media and Wikipedia. The returned format from an API call is often JSON (JavaScript Object Notation) (National Information Standards Organization (NISO), 2016). To sum up, APIs permit the acquisition of scientometric data from various systems via standardized interfaces, which advances research and development in distributed environments and contributes to open bibliometrics as described by Stuart (2018). APIs can also be an important feature to carry out altmetric studies. The field of webometrics gained traction at the beginning of the 2000s, with the analysis of several new online data sources by scientometricians. Webometrics are defined as "the quantitative study of Web-related phenomena" (Thelwall et al., 2005, p. 81). Another milestone in the social media studies of science (Costas, 2018) and the Web in general was the 2010

Altmetrics Manifesto (Priem et al., 2010), while web visibility of scholarly outputs had been previously explored (Aaltojärvi et al., 2008). The goal of the Altmetrics Manifesto was to introduce altmetrics as a supplement to conventional bibliometric indicators due to the increasing amount of scientific information that is available online and to measure how users engage with this data. This development was related to the rise of social media and collaborative online platforms. Regardless of these advances, it is not possible to obtain full open access to scientometric information. For that, the following questions need to be answered, among others.

1. Where can users obtain the data?
2. What restrictions exist to the free re-use of the data for any purpose?
3. How can the data be retrieved automatically?

The openness of scientometric data is a topic that gets more and more attention among the scholarly community (Waltman & Larivière, 2020), for example, regarding the openness of lists of references (Mugnaini et al., 2021; Sugimoto et al., 2018) or the availability of abstracts from publications (Tay et al., 2020)—considering that scientometric research is only possible if access to this research data is guaranteed. As mentioned above, open data sources are needed to facilitate scientometric research. Scientometricians need open data sources to ensure easier accessibility, transparency of results, and replicability of research. This is postulated by many initiatives, for instance, by the founders of the journal *Quantitative Science Studies* (Waltman et al., 2020). Research evaluation by research performing institutions, research funding organisations, research evaluation agencies, researchers and other stakeholders should be based on open data and transparent indicators, according to several international initiatives. These include the San Francisco Declaration on Research Assessment (DORA) (Cagan, 2013), the Leiden Manifesto for Research Metrics (Hicks et al., 2015), and the Hong Kong Principles for assessing researchers (Moher et al., 2020) as well as national initiatives that evaluate this openness (e.g., (Finnish Ministry of Education and Culture, 2016) ). There are several use cases regarding open data sources for scientometric research, of which we present a selection.

1. Comparing the citation counts of the same publication given by multiple bibliometric data sources;
2. matching publications through Digital Object Identifiers (DOIs) between multiple data sources, for example, to enrich bibliographic metadata;
3. estimating the coverage of academic disciplines in different bibliometric data sources;
4. investigating the number of open access publications in different bibliometric data sources; and
5. carrying out scientometric studies, for example, on the reception of research outputs in several data sources.

Our study is part of a larger project on iterative software development based on user friendly design (Hauschke et al., 2018), the ROSI project (see section 1.3).

## 1.2 Related Initiatives

Apart from the ROSI project, the *metrics project released the *Social Media Registry*[2] that was taken into account for this study. While the latter focused only on social media sources, the Registry of Scientometric Data Sources includes other scientometric data sources, such as those that contain bibliometric or patent data. We carried out a literature review of similar registries and the related literature. Registries of data sources and databases are a common practice in academia, for example, to describe national bibliographic databases (Sīle et al., 2017) and services for persistent identifiers (PIDs) in a structured manner (PID Services Registry, 2020). There have been similar initiatives, such as the *metrics project, the EURITO (EU Relevant,

[2] https://metrics-project.net/en/social-media-registry/

Inclusive, Trusted, and Open Indicators for Research and Innovation Policy) project, Data4Impact project, *Open Science Monitor (OSM)* and *Metrics Toolkit*[3].

The *metrics project focused in particular on social media metrics for the social sciences and included user studies of researchers (Lemke et al., 2019). One project outcome was to develop the above-mentioned Social Media Registry. The EURITO consortium carried out an audit of new data sources that could be used as evidence for research and innovation policy (Tippet et al., 2018) and released an overview of pilot cases and potential data sources (EURITO Consortium, 2018). The Data4Impact consortium investigated data sources to develop "a novel end-to-end system for evidence-based, timely and accurate monitoring and evaluation of research and innovation (R&I) activities" (Grypari et al., 2020, p. 22). As part of the development of the Open Science Monitor, data sources were evaluated, and the scholarly community was asked to provide feedback (OSM Consortium, 2018).

The Metrics Toolkit provides an overview of scientometric indicators and related data sources. It is supported by the altmetric data provider Altmetric, Indiana University and the scholarly communications initiative FORCE11. The Metrics Toolkit can be seen as a similar initiative that had been developed before the registry and provides its data under a Creative Commons Attribution (CC BY) 4.0 International License. In contrast to the Metrics Toolkit, we did not form an advisory board because we would like the community to contribute to the data sources. Furthermore, we do not provide detailed information on indicators, for example, journal-based ones, such as the Journal Impact Factor[4]. Our focus lies on itemizing the data sources themselves.

Apart from the Metrics Toolkit, Snowball Metrics is an initiative that provides information on metrics developed by the scholarly community[5]. Furthermore, several EU high-level expert groups discussed data sources as part of their final reports (Cabello Valdes et al., 2017; European Commission, 2018; Guédon et al., 2019; von Schomberg et al., 2019; Wilsdon et al., 2017).

[3] https://www.metrics-toolkit.org/
[4] https://www.metrics-toolkit.org/journal-impact-factor/
[5] https://snowballmetrics.com/

There is also a project that conducted a survey on novel data sources for scholarly communication (Bosman & Kramer, 2016) and an initiative focusing on altmetric data sources based on the input by expert working groups that was conducted by the US National Information Standards Organization (National Information Standards Organization (NISO), 2016). Finally, we took into account the descriptions of data sources by altmetric data providers (Altmetric.com, 2021b; Plum Analytics, 2019).

## 1.3 The ROSI Project

The objective of the ROSI project is to develop a visualization of scientometric indicators based on open data sources while taking into account user needs. We did not develop new indicators, but reused those from openly available data sources. User studies were carried out in qualitative interviews and workshops with focus groups to gather needs and concerns of researchers as potential users of the visualization.

Finally, interviews were conducted to test the usability of the software. As part of the project, we evaluated data sources that are suitable to generate scientometric indicators. This evaluation led to the development of the registry. The project is embedded in the funding line "Quantitative Research on the Science Sector" by the German Federal Ministry of Education and Research (BMBF) and the goal is to collaborate with related projects.

Since we favour data sources that provide or intend to develop an API, the majority of the data sources in the Registry of Scientometric Data Sources fulfill this criterion. We evaluated scientometric data sources to support the scientometric community in developing their own infrastructure, and to provide additional information about these open scientometric data sources.

# 2. METHODS

After carrying out a literature review (see sections 1.1 and 1.2), we started with an evaluation of data sources. For that purpose, we defined a metadata schema and data structure (i.e., sources, indicators, categories, entities) for a technical description. We selected data sources based on concrete criteria, such as relevance for international scientometric studies, and planned or already existing APIs, SPARQL Endpoints or other interfaces to retrieve or download data.

To publish the registry, we developed a website that displays the data sources, metadata schema, goals of the registry, and additional information about the project. An early version of the registry was presented at several workshops, conferences, and announced in a blog article (Lilienthal, 2019) and on social media. We discussed the registry with project partners, including the *metrics project and OAUNI project (Open access publication at universities in Germany: Development and influencing factors)[6], members of the scientometric community and research administrators, thereby obtaining valuable user feedback on additional data sources (including non-open data sources) and possible improvements to the registry.

# 3. RESULTS

Our analysis resulted in a publicly available and collaborative list of data sources that can be used to retrieve scientometric information: the Registry of Scientometric Data Sources. In this section we will describe the metadata schema (section 3.1) and the features of the registry, which presents a general overview of the data sources and detailed views of each single one (with edit option) (section 3.2); and provides a technical overview of the data sources and their connections (section 3.3).

## 3.1 Metadata schema

Entries in the registry are structured according to the following metadata schema (Table 1).

[6] https://www.sub.uni-goettingen.de/en/projects-research/project-details/projekt/oauni/

Table 1

*Metadata Schema in the Registry of Scientometric Data Sources*

| **No.** | **Metadata field** | **Definition** |
|---|---|---|
| 1. | Name | The name of the data source |
| 2. | Link | The URL to the main page of the data source |
| 3. | Link to logo | A (persistent) link to an image |
| 4. | Description | A description of the data source itself |
| 5. | Category | A category that describes the data source (collaborative platform, online social network, repository, data aggregator, other) |
| 6. | Entity | An entity that can be described by this data source (work, person, organisation, event, project) |
| 7. | License | The license of the data |
| 8. | Interface | The link to an API or other interface |
| 9. | Type of interface | The type of interface (REST API, HTTP API, HTTPS API, SPARQL Endpoint, other) |
| 10. | Documentation | The link to the documentation of the interface |
| 11. | Data format | The API response data format (none, JSON, XML, other) |
| 12. | Comment | A text area for any form of comment |

*Note*. A data source may also describe multiple entities, but only one entity can be selected for each data source in the registry. Adapted from https://osl.tib.eu/rosi-registry/about.php

## 3.2 Overview and detailed view of data sources

The registry provides an overview of all data sources at the home page of the application with some basic information including the name, a description and a link to the data source (see Figure 3, and also GitHub page[7] for the current status of the data entries).

[7] https://github.com/TIBHannover/rosi-registry

**Figure 1**

*Home Page of the Registry of Scientometric Data Sources*

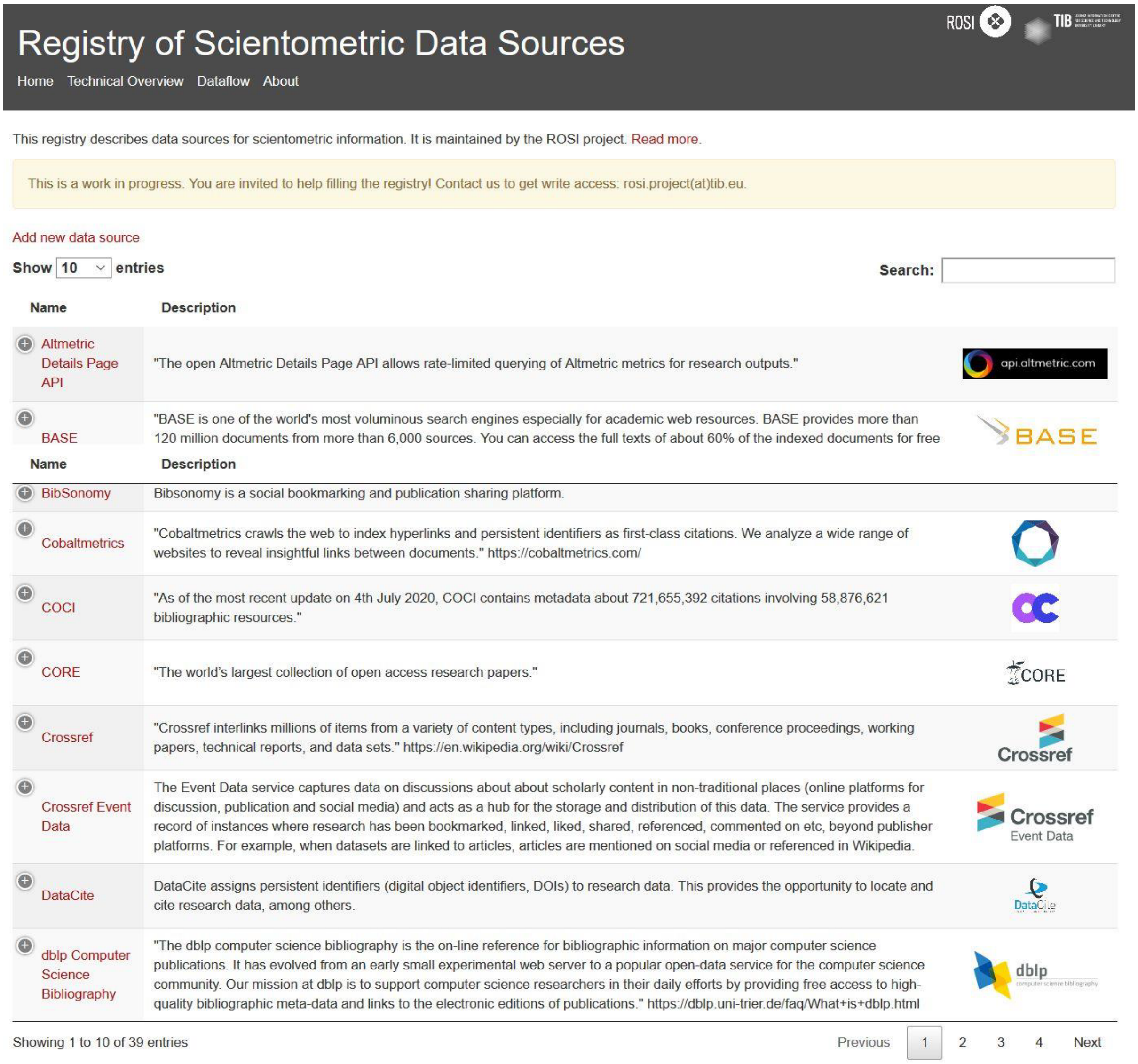
Registry of Scientometric Data Sources

ROSI TIB

Home Technical Overview Dataflow About

This registry describes data sources for scientometric information. It is maintained by the ROSI project. Read more.

This is a work in progress. You are invited to help filling the registry! Contact us to get write access: rosi.project(at)tib.eu.

Add new data source

Show 10 entries Search:

| Name | Description |
|---|---|
| Altmetric Details Page API | "The open Altmetric Details Page API allows rate-limited querying of Altmetric metrics for research outputs." |
| BASE | "BASE is one of the world's most voluminous search engines especially for academic web resources. BASE provides more than 120 million documents from more than 6,000 sources. You can access the full texts of about 60% of the indexed documents for free |
| BibSonomy | Bibsonomy is a social bookmarking and publication sharing platform. |
| Cobaltmetrics | "Cobaltmetrics crawls the web to index hyperlinks and persistent identifiers as first-class citations. We analyze a wide range of websites to reveal insightful links between documents." https://cobaltmetrics.com/ |
| COCI | "As of the most recent update on 4th July 2020, COCI contains metadata about 721,655,392 citations involving 58,876,621 bibliographic resources." |
| CORE | "The world's largest collection of open access research papers." |
| Crossref | "Crossref interlinks millions of items from a variety of content types, including journals, books, conference proceedings, working papers, technical reports, and data sets." https://en.wikipedia.org/wiki/Crossref |
| Crossref Event Data | The Event Data service captures data on discussions about about scholarly content in non-traditional places (online platforms for discussion, publication and social media) and acts as a hub for the storage and distribution of this data. The service provides a record of instances where research has been bookmarked, linked, liked, shared, referenced, commented on etc, beyond publisher platforms. For example, when datasets are linked to articles, articles are mentioned on social media or referenced in Wikipedia. |
| DataCite | DataCite assigns persistent identifiers (digital object identifiers, DOIs) to research data. This provides the opportunity to locate and cite research data, among others. |
| dblp Computer Science Bibliography | "The dblp computer science bibliography is the on-line reference for bibliographic information on major computer science publications. It has evolved from an early small experimental web server to a popular open-data service for the computer science community. Our mission at dblp is to support computer science researchers in their daily efforts by providing free access to high-quality bibliographic meta-data and links to the electronic editions of publications." https://dblp.uni-trier.de/faq/What+is+dblp.html |

Showing 1 to 10 of 39 entries Previous 1 2 3 4 Next

*Note.* The home page displays basic information on the data sources, such as the name, a description, and a link to the data source. The detailed metadata of each data source can be retrieved on a new page of the application. At this view, the registry provides the possibility to search for literature on the data source in the bibliographic databases *CORE (Connecting Repositories)*[8] and *BASE (Bielefeld Academic Search Engine)*[9]. An example of a detailed view of a data source, COCI (the OpenCitations Index of Crossref open DOI-to-DOI citations), is pictured in Figure 2. Screenshot as of September 2020; adapted from https://osl.tib.eu/rosi-registry/index.php

Users are able to enter the metadata of a data source into the registry following an edit link at the detailed view. To track provenance and ensure data integrity, they are required to obtain a password.

[8] https://core.ac.uk/
[9] https://www.base-search.net/

**Figure 2**

*Detailed View of the Data Source, COCI, and its metadata in the Registry of Scientometric Data Sources*

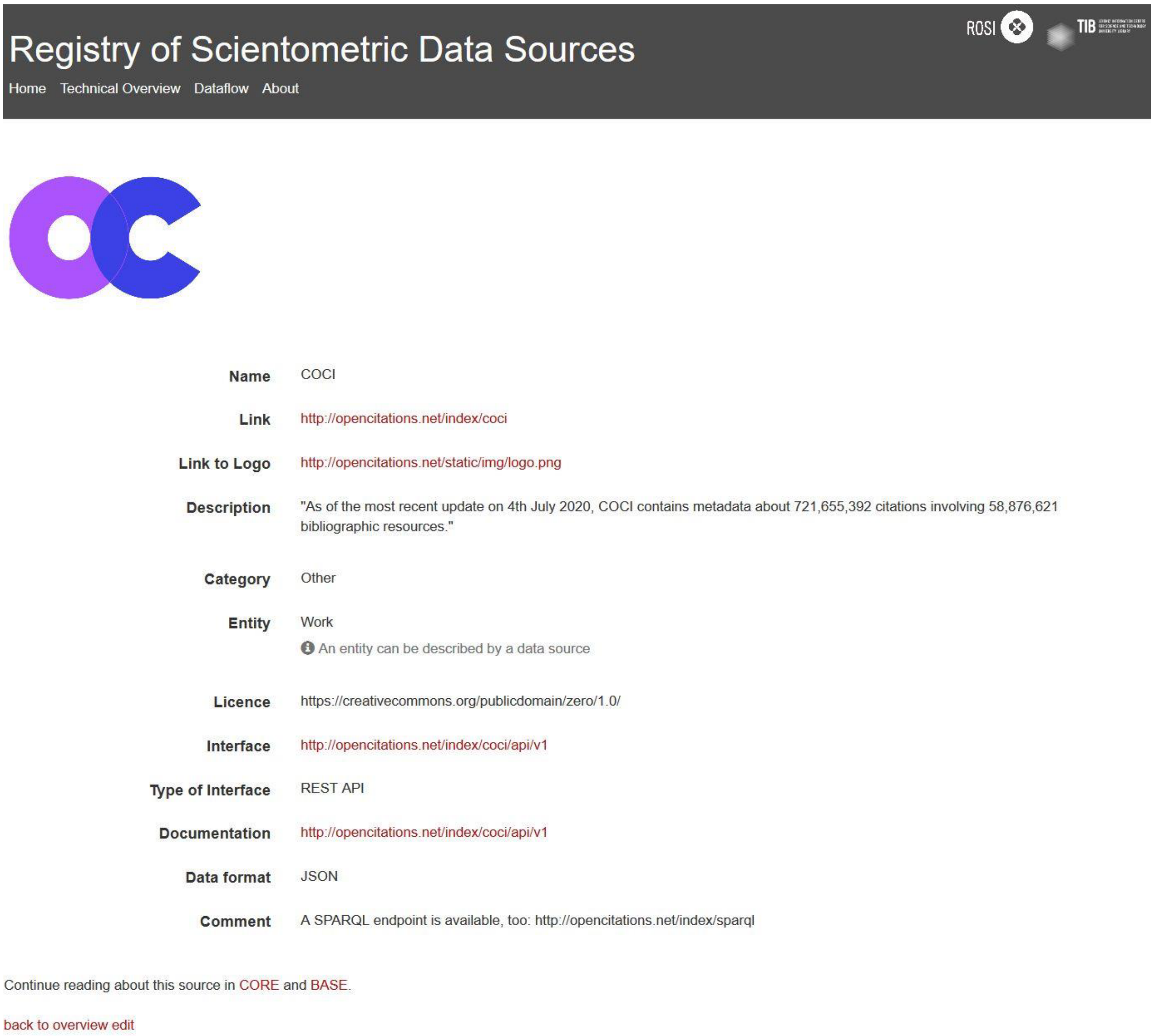

*Note.* Screenshot as of September 2020; adapted from https://osl.tib.eu/rosi-registry/display.php?sourceId=12

## 3.3 Technical overview and network of data sources

In addition to the general overview, a separate page[10] provides a technical overview (see Figure 3) that describes the metadata of each data source. This metadata is useful for third parties who would like to use the data sources for research purposes.

[10] https://labs.tib.eu/rosi/tech.php

**Figure 3**

*Technical Overview of Data Sources in the Registry of Scientometric Data Sources*

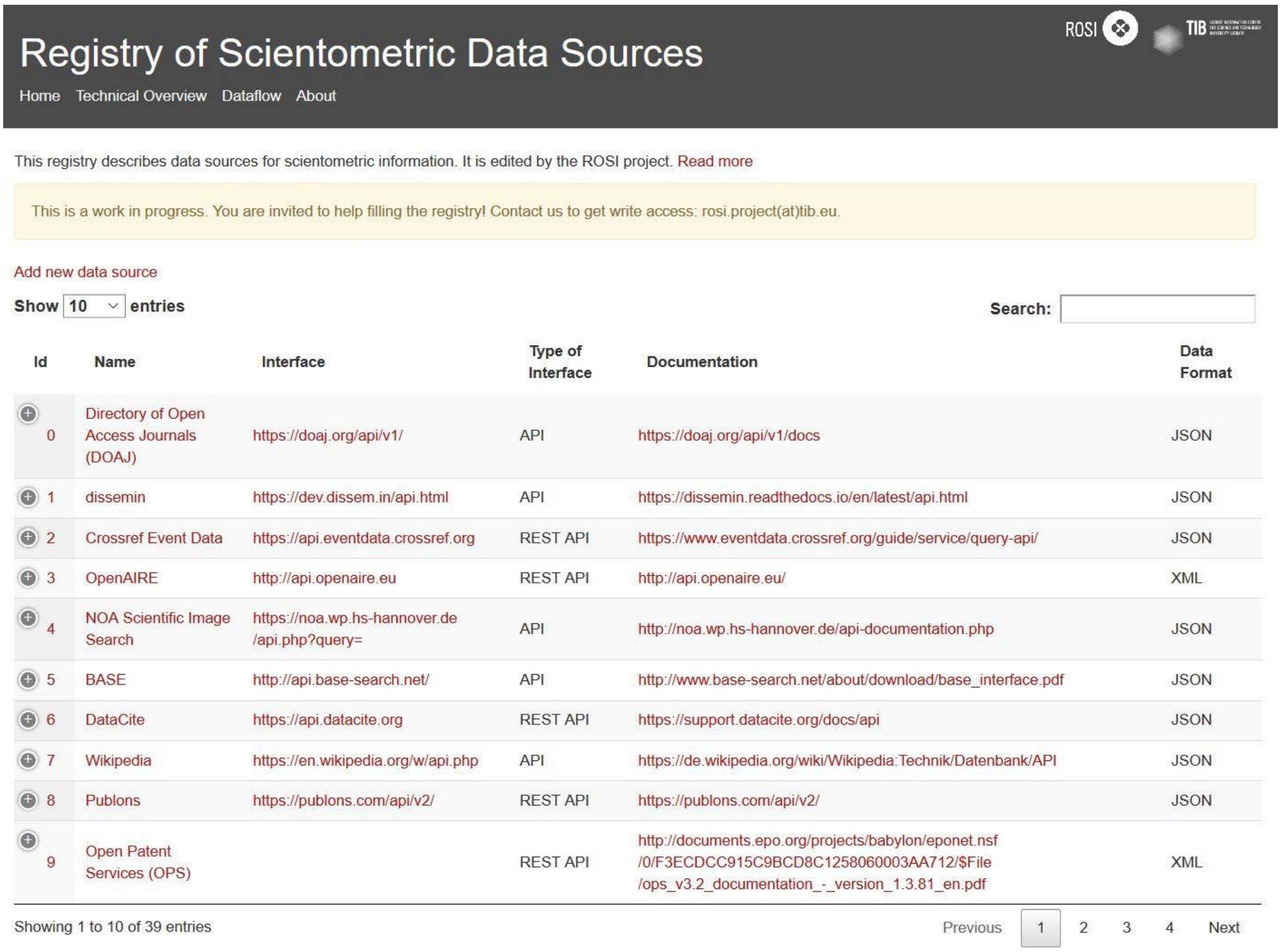
Registry of Scientometric Data Sources

Home Technical Overview Dataflow About

This registry describes data sources for scientometric information. It is edited by the ROSI project. Read more

This is a work in progress. You are invited to help filling the registry! Contact us to get write access: rosi.project(at)tib.eu.

Add new data source

Show 10 entries

Search:

| Id | Name | Interface | Type of Interface | Documentation | Data Format |
|---|---|---|---|---|---|
| 0 | Directory of Open Access Journals (DOAJ) | https://doaj.org/api/v1/ | API | https://doaj.org/api/v1/docs | JSON |
| 1 | dissemin | https://dev.dissem.in/api.html | API | https://dissemin.readthedocs.io/en/latest/api.html | JSON |
| 2 | Crossref Event Data | https://api.eventdata.crossref.org | REST API | https://www.eventdata.crossref.org/guide/service/query-api/ | JSON |
| 3 | OpenAIRE | http://api.openaire.eu | REST API | http://api.openaire.eu/ | XML |
| 4 | NOA Scientific Image Search | https://noa.wp.hs-hannover.de/api.php?query= | API | http://noa.wp.hs-hannover.de/api-documentation.php | JSON |
| 5 | BASE | http://api.base-search.net/ | API | http://www.base-search.net/about/download/base_interface.pdf | JSON |
| 6 | DataCite | https://api.datacite.org | REST API | https://support.datacite.org/docs/api | JSON |
| 7 | Wikipedia | https://en.wikipedia.org/w/api.php | API | https://de.wikipedia.org/wiki/Wikipedia:Technik/Datenbank/API | JSON |
| 8 | Publons | https://publons.com/api/v2/ | REST API | https://publons.com/api/v2/ | JSON |
| 9 | Open Patent Services (OPS) | | REST API | http://documents.epo.org/projects/babylon/eponet.nsf/0/F3ECDCC915C9BCD8C1258060003AA712/$File/ops_v3.2_documentation_-_version_1.3.81_en.pdf | XML |

Showing 1 to 10 of 39 entries

Previous 1 2 3 4 Next

*Note.* Screenshot as of September 2020; adapted from https://osl.tib.eu/rosi-registry/tech.php

To visualize the network between the data sources, we created an undirected data flow graph[11] with the VOSviewer software (van Eck & Waltman, 2010) (Figure 4).

[11] https://labs.tib.eu/rosi/graph.php

**Figure 4**

*Data flow graph as part of the Registry of Scientometric Data Sources*

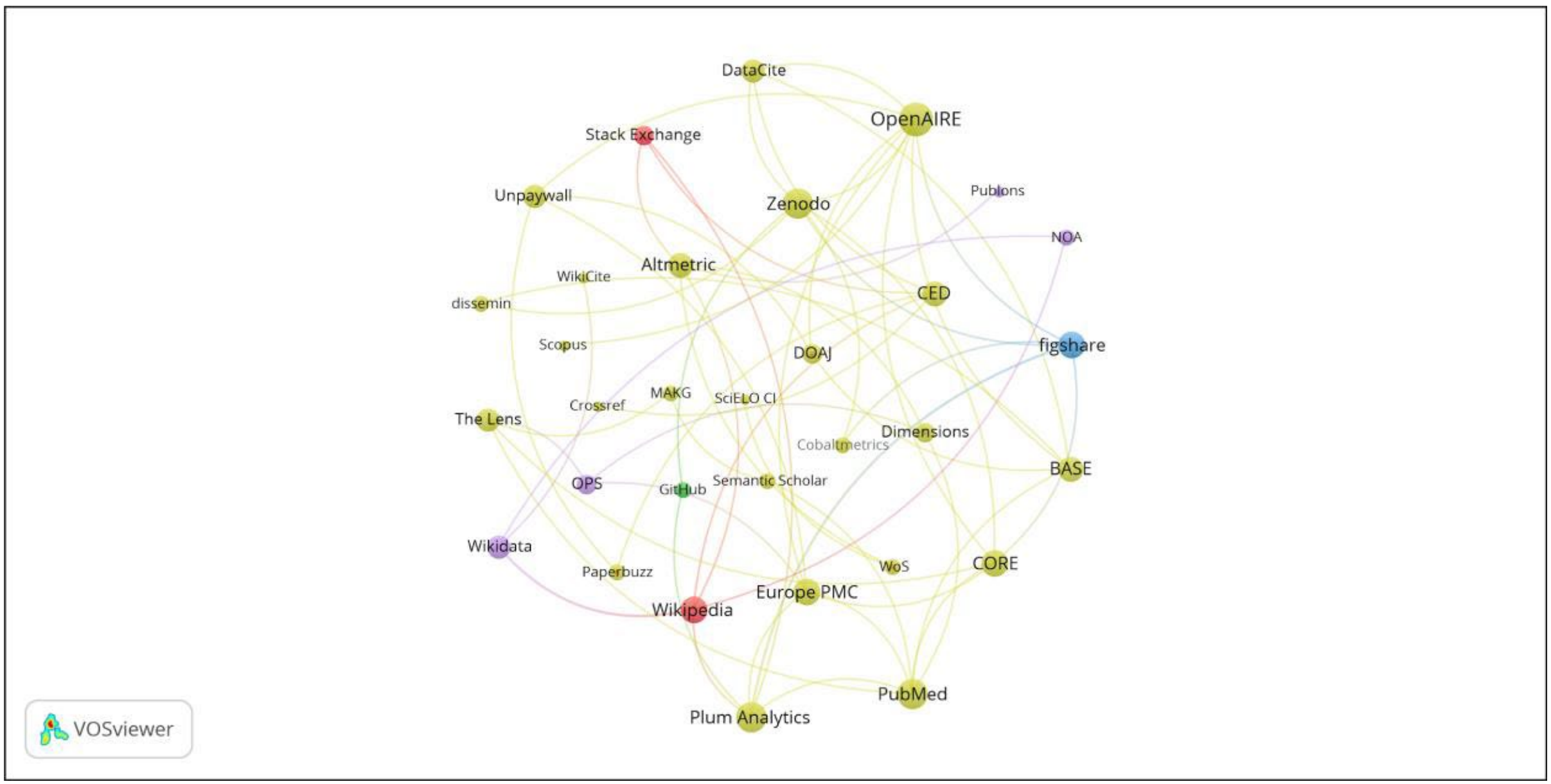

*Note.* The data flow graph illustrates the undirected network between the scientometric data sources (i.e., which source retrieves data from another source or provides data to another source) in the registry. The node size depends on the number of links to other nodes, that is the total link strength. The network may also be opened with the VOSviewer software[12]. Screenshot as of September 2020; adapted from https://osl.tib.eu/rosi-registry/graph.php

# 4. DISCUSSION

Working with the *Registry of Scientometric Data Sources* made it clear that such a collection of data sources can be useful for those with an interest in scientometrics. In the future, more use cases for open data sources are to be included, and we aim to evaluate more data sources. In addition, we would like to extend the metadata schema, which could lead to a data schema as a basis for the Scientometric Data Transfer Protocol (SDTP). The SDTP has been suggested once by Waltman (2019).

Generally speaking, open data sources are constantly changing, and this is why some sources might be added to the registry or replaced in the future. Put simply, "the job is […] never complete" (Wass, 2018, p. 1) and sources need to be constantly monitored and discussed

12 https://www.vosviewer.com/vosviewer.php?map=https://labs.tib.eu/rosi/includes/network.txt=https://labs.tib.eu/rosi/includes/map.txt

(Zahedi & Costas, 2018). To date, we only provide information about a selection of data sources, and we issue a call to the scientometric community and other stakeholders to contribute to the registry. Contributions may range, for instance, from suggesting additional data sources and commenting on existing data sources to extending the metadata schema. Furthermore, a comparison of the coverage of data sources (e.g., (Visser et al., 2021) is currently out of scope, but would be a beneficial feature of the registry. Coming back to our initial research question, namely *Which scientometric data sources are available for research purposes?*, we have described a process for building such a registry which permits feedback and contributions of new information from the scholarly community. We have already received feedback on the registry by some users, but need to find users who are willing to contribute information about new data sources. Since online data sources are dynamic with constant updates and occasional changes in URLs, timeliness and sustainability of such a registry would be more easily guaranteed with a broader involvement of the scientometric community. Low participation rates by other users can also be observed in other calls for participation, for example, to provide citation data to populate open citation data indexes (Heibi et al., 2019). Nevertheless, we expect that the registry can serve the scholarly community in conducting scientometric research and become a resource for users who would like to find out more about scientometric data sources. Ultimately, this approach should contribute to open bibliometrics and scientometrics by providing comprehensive information on data sources and how they are interconnected.

# AUTHOR CONTRIBUTIONS

Conceptualization: Grischa Fraumann, Svantje Lilienthal, Christian Hauschke

Data curation: Grischa Fraumann, Svantje Lilienthal, Christian Hauschke

Formal analysis: Grischa Fraumann, Svantje Lilienthal, Christian Hauschke

Funding acquisition: Christian Hauschke

Investigation: Grischa Fraumann, Svantje Lilienthal, Christian Hauschke

Methodology: Grischa Fraumann, Svantje Lilienthal, Christian Hauschke

Project administration: Christian Hauschke

Resources: Christian Hauschke

Software: Svantje Lilienthal

Supervision: Christian Hauschke

Validation: Grischa Fraumann, Svantje Lilienthal, Christian Hauschke

Visualization: Svantje Lilienthal (RSDS), Grischa Fraumann (data flow graph)

Writing – original draft: Grischa Fraumann

Writing – review & editing: Grischa Fraumann, Svantje Lilienthal, Christian Hauschke

# COMPETING INTERESTS

The authors declare that they have no competing interests.

# FUNDING INFORMATION

This research was funded by the German Federal Ministry of Education and Research (BMBF) under grant number 01PU17019.

# DATA AVAILABILITY

The data is publicly available and can be retrieved from the following URL:

https://github.com/TIBHannover/rosi-registry

# ACKNOWLEDGEMENTS

We started creating the first table with data sources in September 2018, and since then several colleagues and users contributed to this study to develop the registry. We would like to thank the users of the registry, and Ina Blümel, Lambert Heller, Anne Hobert, Gábor Kismihók, Nick Haupka, Katrin Leinweber, Astrid Orth, Cäcilia Schröer, Lucia Sohmen, Graham Triggs, Tatiana Walther, and the students of Information Studies at the University of Copenhagen for feedback on the data sources. We would like to thank Manuel Prinz and Nees Jan van Eck for support on the data flow graph. We would like to thank Raf Guns, Olesya Gladushyna, David Shotton, and Lucia Sohmen for reading and commenting on early drafts of this article. Finally, we would like to thank the participants of the 2019 Open Scientometric Data Infrastructure Workshop at Leiden University, the 2019 *Metrics in Transition Workshop, the 1st OPERA workshop 2019, the 2019 Conference for Research Software Engineers in Germany, and the 2020 BMBF Conference on Quantitative Research on the Science Sector for comments on the registry.

# REFERENCES

Aaltojärvi, I., Arminen, I., Auranen, O., & Pasanen, H.-M. (2008). Scientific Productivity, Web Visibility and Citation Patterns in Sixteen Nordic Sociology Departments. *Acta Sociologica*, *51*(1), 5–22. https://doi.org/10.1177/0001699307086815

Altmetric.com. (2021a). *Attention Sources coverage dates* https://help.altmetric.com/support/solutions/articles/6000240455-attention-sources-coverage-dates

Altmetric.com. (2021b). *Sources of Attention*. https://help.altmetric.com/support/solutions/folders/6000237990

Baas, J., Schotten, M., Plume, A., Côté, G., & Karimi, R. (2020). Scopus as a curated, high-quality bibliometric data source for academic research in quantitative science studies. *Quantitative Science Studies*, *1*(1), 377–386. https://doi.org/10.1162/qss_a_00019

Balstad, M. T., & Berg, T. (2020). A long-term bibliometric analysis of journals influencing management accounting and control research. *Journal of Management Control*, *30*(4), 357–380. https://doi.org/10.1007/s00187-019-00287-8

Birkle, C., Pendlebury, D. A., Schnell, J., & Adams, J. (2020). Web of Science as a data source for research on scientific and scholarly activity. *Quantitative Science Studies*, *1*(1), 363–376. https://doi.org/10.1162/qss_a_00018

Bosman, J., & Kramer, B. (2016). *Innovations in scholarly communication – data of the global 2015–2016 survey.* Zenodo. https://doi.org/10.5281/zenodo.49583

Cabello Valdes, C., Rentier, B., Kaunismaa, E., Metcalfe, J., Esposito, F., McAllister, D., Maas, K., Vandevelde, K., & O'Carroll, C. (2017). *Evaluation of Research Careers fully acknowledging Open Science Practices; Rewards, incentives and/or recognition for researchers practicing Open Science. Working Group on Rewards under Open Science*. https://op.europa.eu/s/y2lT

Cagan, R. (2013). The San Francisco Declaration on Research Assessment. *Disease Models & Mechanisms*, *6*(4), 869–870. https://doi.org/10.1242/dmm.012955

Costas, R. (2018). Towards the social media studies of science: social media metrics, present and future. *arXiv preprint arXiv:1801.04437*. https://arxiv.org/abs/1801.04437

Crossref Event Data. (2021). *Event Data User Guide*. https://www.eventdata.crossref.org/guide/index.html

EURITO Consortium. (2018). *End of Pilot Phase Review*. Zenodo. https://zenodo.org/doi/10.5281/zenodo.3413526

European Commission. (2018). *Open innovation, open science, open to the world: reflections of the Research, Innovation and Science Policy Experts (RISE) High Level Group*. https://op.europa.eu/s/y2lU

European Commission. (2020). *Open access to JRC Research Infrastructures*. https://ec.europa.eu/jrc/en/research-facility/open-access

Finnish Ministry of Education and Culture. (2016). *Evaluation of Openness in the Activities of Research Organisations and Research Funding Organisations in 2016*. http://urn.fi/URN:NBN:fi-fe2016111829246

Garfield, E. (2007). The evolution of the science citation index. *International microbiology*, *10*(1), 65. https://doi.org/10.2436/20.1501.01.10

Gartner (2020). *Application programming interface (API)*. https://www.gartner.com/en/information-technology/glossary/application-programming-interface

Grypari, I., Pappas, D., Manola, N., & Papageorgiou, H. (2020, 11–16 May). Research & Innovation Activities' Impact Assessment: The Data4Impact System. In *Proceedings of the 1st Workshop on Language Technologies for Government and Public Administration (LT4Gov)* (pp. 22–27). https://aclanthology.org/2020.lt4gov-1.4.pdf

Guédon, J.-C., Jubb, M., Kramer, B., Laakso, M., Schmidt, B., Šimukovič, E., Hansen, J., Kiley, R., Kitson, A., van der Stelt, W., Markram, K., & Patterson, M. (2019). *Future of Scholarly Publishing and Scholarly Communication: Report of the Expert Group to the European Commission*. https://op.europa.eu/s/y2lV

Gusenbauer, M. (2019). Google Scholar to overshadow them all? Comparing the sizes of 12 academic search engines and bibliographic databases. *Scientometrics*, *118*(1), 177–214. https://doi.org/10.1007/s11192-018-2958-5

Hauschke, C., Cartellieri, S., & Heller, L. (2018). Reference implementation for open scientometric indicators (ROSI). *Research Ideas and Outcomes*, *4*. https://doi.org/10.3897/rio.4.e31656

Heibi, I., Peroni, S., & Shotton, D. (2019). Crowdsourcing open citations with CROCI--An analysis of the current status of open citations, and a proposal. *arXiv preprint arXiv:1902.02534*. https://arxiv.org/abs/1902.02534

Hicks, D., Wouters, P., Waltman, L., De Rijcke, S., & Rafols, I. (2015). Bibliometrics: the Leiden Manifesto for research metrics. *Nature News*, *520*(7548), 429. https://doi.org/10.1038/520429a

Lariviere, V., & Sugimoto, C. R. (2019). The journal impact factor: A brief history, critique, and discussion of adverse effects. In *Springer handbook of science and technology indicators* (pp. 3–24). Springer.

Lemke, S., Mehrazar, M., Mazarakis, A., & Peters, I. (2019). "When You Use Social Media You Are Not Working": Barriers for the Use of Metrics in Social Sciences. *Frontiers in Research Metrics and Analytics*, *3*(39). https://doi.org/10.3389/frma.2018.00039

Lilienthal, S. (2019). Registry of [Open] Scientometric Data Sources – a collaborative directory of scientometric data sources. *TIB Blog*. https://blogs.tib.eu/wp/tib/2019/05/23/registry-of-open-scientometric-data-sources-a-collaborative-directory-of-scientometric-data-sources/

Lotka, A. J. (1926). The frequency distribution of scientific productivity. *Journal of the Washington Academy of Sciences*, *16*(12), 317–323. http://www.jstor.org/stable/24529203

Moher, D., Bouter, L., Kleinert, S., Glasziou, P., Sham, M. H., Barbour, V., Coriat, A.-M., Foeger, N., & Dirnagl, U. (2020). The Hong Kong Principles for assessing researchers: Fostering research integrity. *PLOS Biology*, *18*(7), e3000737. https://doi.org/10.1371/journal.pbio.3000737

Mugnaini, R., Fraumann, G., Tuesta, E. F., & Packer, A. L. (2021). Openness trends in Brazilian citation data: factors related to the use of DOIs. *Scientometrics*, *126*(3), 2523-2556. https://doi.org/10.1007/s11192-020-03663-7

National Information Standards Organization (NISO). (2016). *NISO Alternative Assessment Metrics (Altmetrics) Initiative*. https://www.niso.org/standards-committees/altmetrics

OSM Consortium. (2018). *Open Science Monitor (OSM). Updated methodological note.* https://research-and-innovation.ec.europa.eu/system/files/2020-01/open_science_monitor_methodological_note_april_2019.pdf

PID Services Registry. (2020). *About*. https://pidservices.org/about

Plum Analytics. (2019). *About PlumX Metrics*. https://plumanalytics.com/learn/about-metrics/

Priem, J., Taraborelli, D., Groth, P., & Neylon, C. (2010). *altmetrics: a manifesto*. http://altmetrics.org/manifesto/

Sīle, L., Guns, R., Sivertsen, G., & Engels, T. C. E. (2017). *European Databases and Repositories for Social Sciences and Humanities Research Output*. https://doi.org/10.6084/m9.figshare.5172322.v2

Stuart, D. (2018). Open bibliometrics and undiscovered public knowledge. *Online Information Review*, *42*(3), 412–418. https://doi.org/10.1108/OIR-07-2017-0209

Sugimoto, C. R., Murray, D. S., & Larivière, V. (2018). Open citations to open science. *ISSI Blog*. https://www.issi-society.org/blog/posts/2018/april/open-citations-to-open-science/

Tay, A., Kramer, B., & Waltman, L. (2020). Why openly available abstracts are important – overview of the current state of affairs. *Leiden Madtrics*. https://leidenmadtrics.nl/articles/why-openly-available-abstracts-are-important-overview-of-the-current-state-of-affairs

Thelwall, M., Vaughan, L., & Björneborn, L. (2005). Webometrics. *Annual review of information science and technology*, *39*(1), 81–135.

Tippet, C., Blind, K., & Parraguez Ruiz, P. (2018). *Audit of new R&I data*. Zenodo. https://zenodo.org/doi/10.5281/zenodo.1404116

van Eck, N. J., & Waltman, L. (2010). Software survey: VOSviewer, a computer program for bibliometric mapping. *Scientometrics*, *84*(2), 523-538. https://doi.org/10.1007/s11192-009-0146-3

Visser, M., van Eck, N. J., & Waltman, L. (2021). Large-scale comparison of bibliographic data sources: Scopus, Web of Science, Dimensions, Crossref, and Microsoft Academic. *Quantitative Science Studies*, 1–22. https://doi.org/10.1162/qss_a_00112

von Schomberg, R., Holbrook, J. B., Oancea, A., Kamerlin, S. C. L., Ràfols, I., Jacob, M., & Wouters, P. (2019). *Indicator frameworks for fostering open knowledge practices in science and scholarship. Report of the Expert Group on Indicators for Researchers' engagement with open science*. https://op.europa.eu/s/y2lR

Waltman, L., & Larivière, V. (2020). Special issue on bibliographic data sources. *Quantitative Science Studies*, *1*(1), 360–362. https://doi.org/10.1162/qss_e_00026

Waltman, L., Larivière, V., Milojević, S., & Sugimoto, C. R. (2020). Opening science: The rebirth of a scholarly journal. *Quantitative Science Studies*, *1*(1), 1–3. https://doi.org/10.1162/qss_e_00025

Wass, J. (2018, 25 May). Five principles for community altmetrics data. *altmetrics18: Science & The Public: Public Interactions with Science through the Lens of Social Media*, London. http://altmetrics.org/wp-content/uploads/2018/04/altmetrics18_paper_4_Wass.pdf

Wilsdon, J., Bar-Ilan, J., Frodeman, R., Lex, E., Peters, I., & Wouters, P. (2017). *Next-generation metrics. Responsible metrics and evaluation for open science: Report of the European Commission Expert Group on Altmetrics*. https://op.europa.eu/s/y2lS

Zahedi, Z., & Costas, R. (2018). General discussion of data quality challenges in social media metrics: Extensive comparison of four major altmetric data aggregators. *PLOS ONE*, *13*(5), e0197326. https://doi.org/10.1371/journal.pone.0197326